\documentclass[twocolumn,prl,amsmath,amssymb]{revtex4}% Physical Review B

\usepackage{graphicx}% Include figure files
\usepackage{dcolumn}% Align table columns on decimal point
\usepackage{bm}% bold math
\begin{document}

\preprint{ECM}

\title{Long range Ising model  for credit risk modeling in homogeneous portfolios}

\author{Jordi Molins}
\affiliation{%
Dept. de Matem\`atica Aplicada I, Universitat Polit\`ecnica de Catalunya \\ Diagonal 647, ETSEIB, 08028 Barcelona, Catalonia (Spain)}

\author{Eduard Vives}
\email{eduard@ecm.ub.es}
\affiliation{%
Dept. d'Estructura i Constituents de la Mat\`eria, Universitat de Barcelona \\
Diagonal 647, Facultat de F\'{\i}sica, 08028 Barcelona, Catalonia (Spain)}

\date{\today}

\begin{abstract}
Within the  framework of  maximum entropy principle  we show  that the
finite-size  long-range Ising  model  is the  adequate  model for  the
description of  homogeneous credit  portfolios and the  computation of
credit  risk  when  default  correlations between  the  borrowers  are
included.   The exact  analysis of  the  model suggest  that when  the
correlation increases a first-order-like transition may occur inducing
a  sudden risk  increase.  Such  a feature  is not  reproduced  by the
standard models used in credit risk modeling.
\end{abstract}

\maketitle

The  modelization  and  control  of  risk  is  a  subject  of  extreme
importance in  a rich variety  of disciplines ranging from  geology to
medicine. To know which is the probability distribution of rare events
that may cause very important losses is the key point to be solved and
to which a lot of efforts have been devoted for many years. Within the
framework of  finance, the  most studied case  is that of  market risk
(the risk  associated to uncertainties  of capital markets).   But, by
far,  the  risk that  most  concern  practitioners  (both due  to  its
negative  impact  and  its   modeling  difficulties)  is  credit  risk
\cite{Bielecki2002}: a  bank or financial institution  give credits to
$N$ borrowers which may fail to  return it.  Such a group of borrowers
constitutes a  so called  credit portfolio.  It  is very  important to
model which is the probability $p(L)$  that $L$ ($0 \leq L \leq N$) of
such borrowers  will fail  in returning the  credit. The  knowledge of
$p(L)$ is of extreme importance  to decide the total amount of capital
and reserves  that the bank should  hold to prevent a  failure and for
economic and regulatory reasons. Also $p(L)$ is the main driver in the
valuation and hedging of credit derivatives.

Banks have  been using many alternative  models.  Most of  these, as a
first step, consider the case  of homogeneous portfolios for which the
properties of the  borrowers and their credits are  supposed to be the
same. In this paper we will also restrict to this homogeneous case. In
the  financial literature  the description  of a  portfolio  starts by
defining the  ``losses'' $l_i$ ($i=1,\dots,N$) which  takes values $0$
and $1$  depending on whether the  borrower $i$ returns  the credit or
fails to  return it.  For  convenience, we define $S_i=1-2  l_i$ which
takes values $+1$ when a borrower  returns the credit and $-1$ when it
fails to return it.  Note that the number of losses is $L=\sum_i l_i =
(N-\sum_i S_i)/2$.

The  simplest model  is  the  one of  independent  borrowers. This  is
characterized  by  a single  parameter  which  is  the probability  of
default  $p$. In  this  case, the  loss  probability is  given by  the
binomial distribution:
\begin{equation}
p_B(L)=  \left  (  \begin{array}{c}   N  \\  L  \end{array}  \right  )
(1-p)^{N-L} p^{L}.
\label{binomial}
\end{equation}
The expected value  of the number of losses is $\langle  L \rangle = N
\langle l_i \rangle = N p$  and its variance is $V(L)= N^2 V(l_i)= N^2
p(1-p)$.   (The  corresponding  expressions  in  terms  of  the  $S_i$
variables can be easily derived  taking into account that $\langle S_i
\rangle = N(1-2p)$ and $V(S_i)=  4p(1-p)$.)  The main drawback of this
model is  that it  does not contain  the possibility  for correlations
between borrowers which are known to exist and play an important role.
Although in some cases the links between borrowers can be difficult to
understand, in many  cases borrowers tend to default  in a cooperative
manner.

The definition of the default correlation $\rho$ in terms of the $S_i$
variables is given by:
%
%\begin{eqnarray}
%\label{rho}
%\rho & = & \frac{\langle l_i  l_j \rangle - \langle l_i  \rangle \langle l_j
%\rangle    }{\sqrt{\langle    l_i^2     \rangle    -    \langle    l_i
%\rangle^2}\sqrt{\langle     l_j^2     \rangle     -    \langle     l_j
%\rangle^2}} \nonumber \\ & = & \frac{\langle  S_i  S_j \rangle  -  \langle S_i%\rangle \langle S_i  \rangle}
%{\sqrt{\langle S_i^2 \rangle - \langle S_i \rangle^2} \sqrt{\langle S_j^2 
%\rangle - \langle S_j \rangle^2}   } \; \; \; (i\neq j).
%\end{eqnarray}
%
\begin{equation}
\label{rho}
\rho = \frac{\langle S_i S_j \rangle - \langle S_i \rangle \langle S_j
\rangle  }{\sqrt{\langle  S_i^2   \rangle  -  \langle  S_i  \rangle^2}
\sqrt{\langle S_j^2 \rangle - \langle S_j \rangle^2} } \; \; \; (i\neq
j).
\end{equation}
The same  formula applies for the  $l_i$ variables, as  can be checked
straightforwardly.   To  include  correlations  different  theoretical
models  have  been proposed  by  modifying  the  model of  independent
borrowers including phenomenological hypothesis  on how $p$ depends on
external  factors  \cite{Martin2001}.    For  instance,  the  economic
conditions can be described by a latent variable $y$ which is a random
variable with a certain probability density $f(y)$.  Then, the concept
of  default   probability  $p$  is  substituted  by   the  concept  of
conditional default probability $\phi(y)$  such that for a fixed value
of   $y$  the   probability   of  $L$   defaults   is  binomial   with
$p=\phi(y)$. Therefore,
\begin{equation}
p(L)= \int_{-\infty}^{\infty} dy \, f(y) \; \left  (  \begin{array}{c}   N  \\  L  \end{array}  \right  )
\left [ 1-\phi(y) \right ]^{N-L} \phi(y)^{L}.
\end{equation}
Such   conditionally   independent    models   (CIM)   are   typically
characterized by two parameters which could be empirically determined.
First,  the average  (also called  unconditional)  default probability
given by:
\begin{equation}
{\bar  p}  \equiv \langle  l_i  \rangle  = \int_{-\infty}^{\infty}  dy
\phi(y) f(y).
\end{equation}
Second,  the default correlation  which, for  the case  of homogeneous
portfolios, is given by:
\begin{equation}
\rho=   \frac{\int_{-\infty}^{\infty}  \phi(y)^2   f(y)  dy   -  \left
[\int_{-\infty}^{\infty}       \phi(y)       f(y)      dy       \right
]^2}{\int_{-\infty}^{\infty}     \phi(y)    f(y)     dy     -    \left
[\int_{-\infty}^{\infty} \phi(y) f(y) dy \right ]^2}.
\end{equation}
For  instance,  the Merton  based  model  (used  by JPMorgan  and  KMV
companies  in  their  credit  portfolio  models)  corresponds  to  the
assumption that  $f(y)$ is  the $N(0,1)$ Gaussian  probability density
and the function $\phi$ is:
\begin{equation}
\label{phim} \phi_M(y)       = \Phi \left(
\frac{c-\sqrt{q}y}{\sqrt{1-q}} \right)\equiv
\int_{-\infty}^{\frac{c-\sqrt{q}y}{\sqrt{1-q}}}
\frac{1}{\sqrt{2\pi}} e^{-z^2/2} dz,
\end{equation}
where  $\Phi$ is the error function, $c$  is   the ``critical
value'' and $q$ is the ``asset correlation''  (which should  not
be confused with the default correlation).   For this case  it is
easy to check that the average default probability  is ${\bar p} =
\Phi(c)$, and that  the default correlation is:
\begin{equation}
\rho =   \int_{-\infty }^c  \int_{-\infty}^c   \frac{du  \;   dv}
{2\pi\sqrt{1-q^2}} \; e^{-(u^2+v^2-2quv)/2(1-q^2)}.
\end{equation}
Other examples of CIM  are the CreditPortfolioView model (of McKinsey)
and  the CreditRisk$+$  model  (of Credit  Suisse Financial  Products)
which    use     different    functions    $\phi(y)$     and    $f(y)$
\cite{Crouhy2000,Gordy2000}.  The CIM have  been shown to only capture
cyclic correlations, leaving  aside direct correlations, contagion and
cascade effects \cite{Gieseke2002}.

In  this  letter we  propose  a different  method  to  model a  credit
portfolio  that  includes  default  correlations  in  a  fundamentally
different  way.   We  describe  portfolios with  vectors  $X=(S_1,S_2,
\cdots,  S_N)$.   The  set  of  all possible  $2^N$  vectors  $X$  is,
therefore,  our  phase  space   $\Omega=\{X\}$.   To  solve  a  credit
portfolio model  is to  assign a probability  law $P(X)$  on $\Omega$.
Instead of  trying to  derive which are  the dependencies  between the
different borrowers or between the borrowers and the external factors,
we will assume that default  correlation $\rho$ is different from zero
and  derive $P(X)$  from  the Maximum  Statistical Entropy  principle.
This  principle, which is  essential in  Information Theory,  has been
widely used  for building probability distributions  when a underlying
theory  is lacking.   It states  that  one should  consider the  model
$P(X)$ that maximizes the entropy functional:
\begin{equation}
S[ P] = \sum_{X \in \Omega} P(X) \ln P(X),
\label{entropy}
\end{equation}
subject to  the conditions imposed by the  previous known information.
Such a method allows to control what are the exact ingredients (or the
exact  information)  that  is  taken  into account  to  formulate  the
model.  This is  opposite to  the above  presented  theoretical models
which assume  a rigid structure  for relations between  borrowers that
cannot be empirically checked.

It  is instructive  to first  present the  derivation of  the simplest
model.  We  will impose only  two conditions. First  the normalization
condition:
\begin{equation}
\sum_{X \in \Omega} P(X) = 1.
\label{cond1}
\end{equation}
Second we  assume that  we have a  previous knowledge of  the expected
value of the number of losses $\langle L \rangle$ or, equivalently, of
the average default probability $\bar p$.  In terms of the probability
law $P(X)$ this condition is written as:
\begin{equation}
\langle \sum S_i \rangle = \sum_{ X \in \Omega} P(X) M(X) = N(1-2{\bar
p}),
\label{cond2}
\end{equation}
where  we have  defined the  function $M(X)=\sum_i  S_i$.  We  want to
stress that the knowledge of  the average default probability $\bar p$
allows  to fix  the expected  value of  $M$ but  not its  exact value.
Maximize    (\ref{entropy})   with   conditions    (\ref{cond1})   and
(\ref{cond2}) is equivalent to maximize the Lagrange functional:
\begin{eqnarray}
\Psi_0[P] &=& \sum_{X \in \Omega} P(X) \ln P(X) - \lambda \left ( \sum_{X \in
\Omega} P(X)  - 1 \right  ) \nonumber \\
&-& \alpha  \left ( \sum_{X \in  \Omega} M(X)
P(X) - N(1-2{\bar p}) \right ),
\end{eqnarray}
where $\lambda$ and $\alpha$  are Lagrange multipliers.  The result of
the maximization is $P(X)  = e^{\lambda-1+\alpha M(X)}$. The parameter
$\lambda$ can be easily deduced imposing (\ref{cond1}). 
%
%This leads to:
%
%\begin{equation}
%P(X) = \frac{e^{\alpha M(X)}}{\sum_{X \in \Omega} e^{\alpha M(X)}}=
%\frac{e^{\alpha M(X)}}{(e^\alpha + e^{-\alpha})^N}.
%\label{px}
%\end{equation}
%
To obtain $\alpha$ we require $\langle M(X) \rangle = N(1-2{\bar p})$.
A straightforward algebra renders $\alpha = \tanh^{-1} (1-2 \bar{p})$.
The probability $P(X)$ results in:
\begin{equation}
P(X)=  (1-{\bar p})^{\frac{N+M(X)}{2}} {\bar p}^{\frac{N-M(X)}{2}}.
\label{binomial2}
\end{equation}
Note that  this probability depends  on $X$ only through  the function
$M(X)$. Since there  are $\tbinom{N}{L}$  different  vectors  that  give  the  same  loss  $L$,  the
probability $p(L)$  turns out to be  Eq. (\ref{binomial}) substituting
$p$ by $\bar p$.

The inclusion of the  existence of correlations between borrowers
into the model requires to introduce  a third condition in the
maximization of (\ref{entropy}). In this case,  the symmetric
function of $X$ whose expected value is fixed by the default
correlation $\rho$ is:
\begin{equation}
H(X)=\frac{2}{N-1}\sum_{i>j} S_i S_j.
\end{equation}
Note that  the number of  terms in the  sum is $N(N-1)/2$ so  that
the pre-factor  $2/(N-1)$   ensures  that  $H(X)  \propto   N$.
Using  (\ref{rho}), (\ref{cond2})  and  the  fact   that $\langle
S_i^2 \rangle=1$ it is  straigforward to check that the knowledge
of $\rho$ allows to  fix the average  value of $\langle H(X)
\rangle$ according to:
\begin{equation}
\langle H(X) \rangle=N \left [ 1-4{\bar p}(1-{\bar p})(1-\rho) \right ].
\label{cond3}
\end{equation}
Imposing  this   condition, together   with conditions
(\ref{cond1}) and  (\ref{cond2}),  leads  to the following
Lagrange functional:
\begin{equation}
\Psi[P] = \Psi_0[P]- \beta \left [ \sum_{X \in \Omega} H(X) P(X)
- \langle H(X) \rangle \right ].
\end{equation}
Diferentiating and  imposing (\ref{cond1}) to  eliminate $\lambda$ one
obtains:
\begin{equation}
P(X)=\frac{e^{\alpha M(X)+\beta H(X)}}{ \sum_{X \in \Omega} e^{\alpha M(X)+\beta H(X)}}.
\label{isi1}
\end{equation}
This can be be rewriten as $P(X)=e^{-\cal H}/ \sum_{X \in \Omega} e^{-
\cal H}$ with:
\begin{equation}
{\cal H} (X) = -\frac{2 \beta}{N-1} \sum_{i<j} S_i S_j - \alpha \sum_{i=1}^{N} S_i,
\end{equation}
which is  the canonical probability distribution  corresponding to the
well known long-range  Ising model (LRIM) of a  finite system with $N$
spins and a  convenient redefinition of the exchange  constant and the
external field  \cite{Tsypin2000}.  Note that  $P(X)$ in (\ref{isi1}),
again, is only a function  of $M(X)$ since $H(X) = \frac{1}{N-1} \left
( M(X)^2 - N \right )$.   Therefore the probability of $L$ losses will
be  obtained by  multiplying expression  (\ref{isi1}) by  the binomial
number $\tbinom{N}{L}$.

To determine the  values of $\alpha$ and $\beta$  in (\ref{isi1}) that
allow to  impose conditions (\ref{cond2}) and  (\ref{cond3}) for given
values  of $\bar  p$ and  $\rho$  is a  difficult task  that has  been
extensively studied  in the field of Statistical  Mechanics. Note that
the  problem when $N  \rightarrow \infty$  can be  easily solved  by a
saddle point  method.  This method  renders a solution  that coincides
with   the   solution   obtained   by  the   mean   field   hypothesis
\cite{Stanley1971}.  But, such an  approximation is of little interest
here because of two  reasons: credit portfolios, although being large,
are  quite far  from the  thermodynamic limit  and such  a  mean field
approximation   inmediately  leads   to  $\rho=0$   (except   for  the
unrealistic case of ${\bar p}=0.5$).

%For a  finite system an  exact numerical calculation, summing  up
%over the  whole $\Omega$  space, can  be performed  up to  $N
%\simeq 500$. The limitation comes from the maximum number that can
%be stored  in  a standard double precision variable. Slightly
%larger values  of $N$  can be analyzed by ad-hoc quadruple  or
%even larger precision codes. For much larger values $N  > 1000$
%approximate solutions would require MC simulations.

\begin{center}
\begin{figure}[th]
\includegraphics[width=7.5cm,clip]{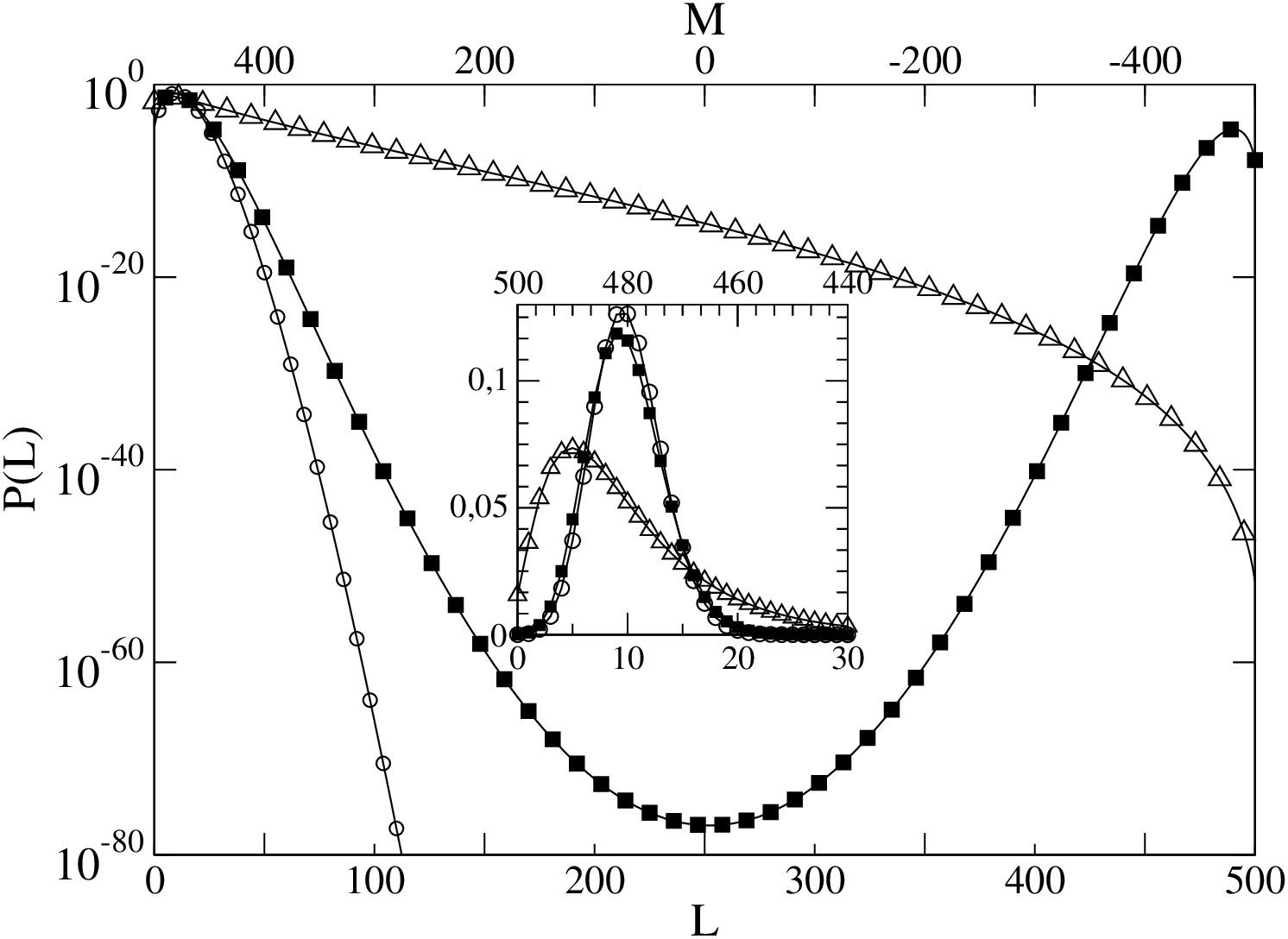}
\caption{\label{FIG1}  Comparison of the  probability of  $L$ failures
$p(L)$ (in linear-log scale)  obtained from the LRIM ($\blacksquare$),
the   binomial  model   ($\bigcirc$)  and   the  Merton   based  model
($\triangle$) with $N=500$ and ${\bar  p}=0.02$. Both the LRIM and the
Merton model correspond  to $\rho=0.01$.  The inset shows  the peak on
the left  in linear scale.  The  scales on top  show the corresponding
values of $M$.}
\end{figure}
\end{center}

Fig.~\ref{FIG1} shows an example  of the probability $p(L)$ derived by
the   LRIM   corresponding   to   ${\bar  p}=0.02$   and   $\rho=0.01$
($\alpha=8.85  \;   10^{-3}$,  $\beta=1.01148$).   The   behaviour  is
compared with the binomial  model (without correlation) and the Merton
based  model.   In  this  last  case the  parameters  ($c=-2.054$  and
$q=0.0718$) have been  calibrated in order to have  the same values of
${\bar p}$ and $\rho=0.01$. Note that  the LRIM is the only model able
to capture  the existence  of a  second peak in  the large  $L$ region
(negative  $M$   region)  that  appears   due  to  the   existence  of
correlations which favour a collective behaviour of the borrowers.

\begin{center}
\begin{figure}[th]
\includegraphics[width=6.8cm,clip]{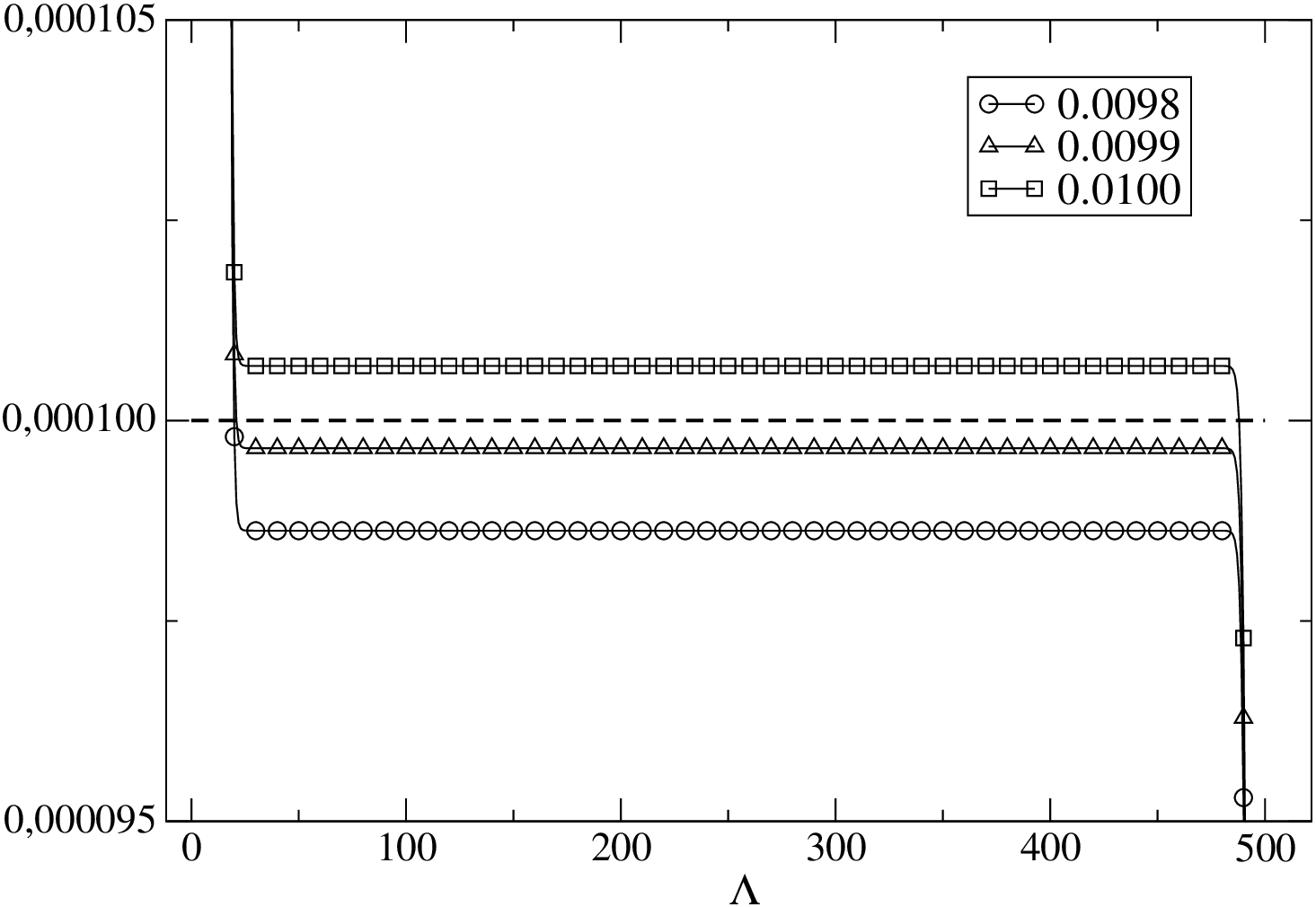}
\caption{\label{FIG2} Calculation  of the  VaR for a  given confidence
level  $\alpha_c=0.0001$  (dashed line).   The  continuous lines  with
symbols  show  the  behaviour  of  the  left  hand  side  in  equation
\ref{lambda1}  as a  function of  $\Lambda$ for  the LRIM  with ${\bar
p}=0.01$ and different values of $\rho$, as indicated by the legend.}
\end{figure}
\end{center}

\begin{center}
\begin{figure}[th]
\includegraphics[width=7.7cm,clip]{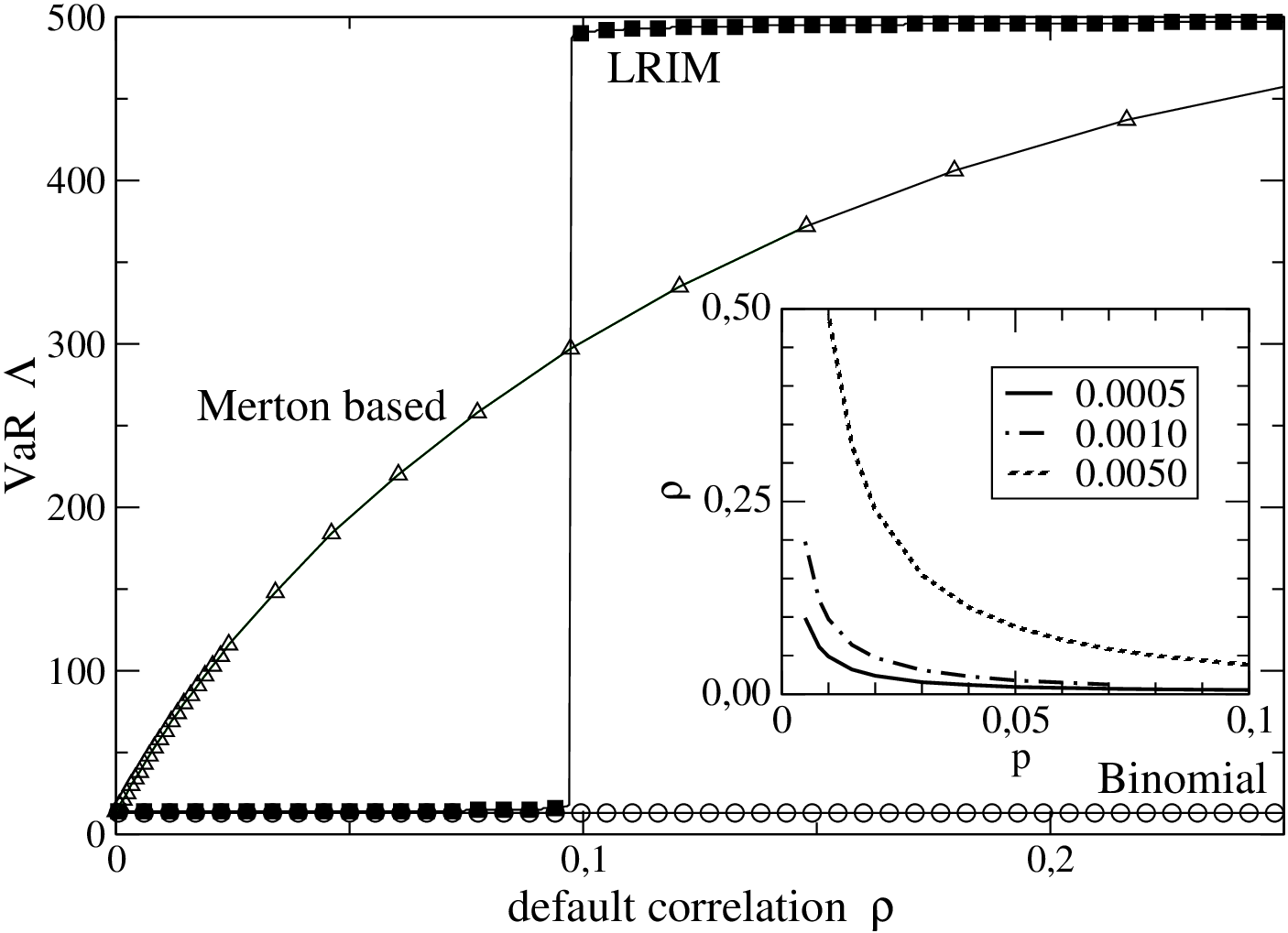}
\caption{\label{FIG3}  VaR as  a function  of the  default correlation
$\rho$  for  the  LRIM,  the  Merton  model  and  the  binomial  model
(independent of  $\rho$). The three  models correspond to  $N=500$ and
$\bar p =  0.01$. The inset shows the position  of the sharp increases
of the  VaR in the  $\bar p$-$\rho$ space  predicted by the  LRIM, for
different values  of the confidence  level $\alpha_c$, as  indicated by
the legend. }
\end{figure}
\end{center}

Although the peak in the right(large $L$ region) is usually very small
compared to  the peak  in the  left (small $L$  region) it  has strong
implications  in the evaluation  of credit  risk. There  are different
strategies to measure risk. A common  one is the computation of the so
called Value-at-Risk  (VaR): this is the number  of failures $\Lambda$
that must  be considered  in order to  ensure that the  probability of
having  more than  $\Lambda$  fails  is smaller  than  a certain  tiny
confidence level $\alpha_c$.  Thus  $\Lambda$ is given by the solution
of the equation:
\begin{equation}
\sum_{L=\Lambda}^N p(L)= \alpha_c
\label{lambda1}
\end{equation}
The   smaller   the    confidence   level,   the   larger   $\Lambda$.
Fig.~\ref{FIG2} shows  an example of the  behaviour of the  sum in the
left of  (\ref{lambda1}) as a  function of a variable  $\Lambda$.  The
horizontal line  indicates the confidence level.  Note  that given the
two-peaked  shape of $p(L)$,  the sum  in (\ref{lambda1})  exhibitis a
constant plateau. Consequently a small  change in $\rho$ may produce a
drastic change in the position of the VaR (crossing point).
From the  physical point  of view,  this change is  very similar
to a first-order  phase  transition  in  which the  order
parameter  (VaR) exhibits a  discontinuity. The main  difference
is that our  system is small and therefore real discontinuities do
not exhist but, instead, there are sharp changes.
In  order  to  make clear  this  sharp  behaviour  we show  in  Figure
\ref{FIG3} the  VaR as  a function of  $\rho$ for ${\bar  p}=0.01$ and
$\alpha_c=0.0001$.  The  Merton model  displays an smooth  increase of
the  VAR  with  increasing  correlation. For  small  correlations  the
differences  between the  LRIM and  the binomial  one  are negligible.
However, there exist a value  of the correlations for which, according
to  the LRIM,  a dramatic  increase in  the VaR  occurs, which  is not
reproduced by the Merton model.
\vspace{-5mm}
It is of  paramount importance to locate the  region where such strong
change in the VaR occurs  for different values of the confidence level
$\alpha_c$.   This is  shown  in the  inset  of Fig.~\ref{FIG3}.   The
linear axes represent the space of empirical parameters ${\bar p}$ and
$\rho$.   The different lines,  corresponding to  different confidence
levels,  separate the region  in which  the behaviour  of the  LRIM is
similar to the binomial model (below) from the region where collective
failure  may  occur  and  thus  predictions with  the  binomial  model
strongly uderstimate  the VaR.  Other  measurements of risk  (like the
expected shortfall) may exhibit even stronger diferences.

There are  a series of possible  extensions of the  LRIM model: first,
the   extension  to   the  case   of  non-homogeneous   portfolios  is
straightforward.   The borrowers  need  to be  separated in  different
tranches. For each  tranche the formulation of the  model will require
the knowledge  of the default probability,  the autocorrelation within
each tranche and the  correlations among different tranches. Note that
in  the LRIM  model will  be easy  to introduce  negative correlations
between  certain  tranches  to  represent  big  competitors  in  small
markets.   This  is  impossible  to  reproduce  with  the  conditional
independent models.  A second  interesting extension will be to change
the discrete  variables $S_i$  to continuous variables  $\phi_i$.  The
Ising  model, then,  be substituted  by a  $\phi^4$ model.   This will
enable to  describe the  fact that when  a borrower defaults  there is
always a recovery of part of the loan.  A third extension could be the
inclusion  of randomness  in  the LRIM.   For  instance random  fields
and/or random bonds would allow to take into account the uncertainties
in  the knowledge  of the  empirical values  of $\rho$  and  $\bar p$.
Finally one  could include higher order interaction  terms, like three
spin terms.   This would allow to  introduce, in a  controled way, the
existence of three borrower correlations  which are know to exist when
loans  are   fully  guaranteed  by  third  parties.    Note  that  the
conditional  independent   models  do  also   introduce  higher  order
correlations, but in a completely uncontrolled manner.

Within the framework of information theory, it has been shown that the
LRIM is  the natural model  to be used  for the description  of credit
portfolios and  risk modelling when  knowledge about the  existence of
correlations is taken  into account.  One of the  main results is that
risk measurements can  display big and sharp increases  due to default
correlations, in  regions of the  space parameters where  other models
fail to  predict them.  The  discussion presented also shows  that the
LRIM  can be  useful not  only  in finance  but, in  general, for  any
complex system  described by binary variables for  which the available
information  is   the  average   value  and  the   correlations. 

J.M.  acknowledges  fruitful discussions with  Josep Masdemont.  E.V.
acknowledges   fruitful  discussions   with   A.Planes,  J.Vives   and
R.S.Johal.   This  work  has  received financial  support  from  CICyT
(Spain),   project  MAT2001-3251   and   CIRIT  (Catalonia),   project
2000SGR00025.

\end{document}